\documentclass[conference]{IEEEtran}
\IEEEoverridecommandlockouts
\usepackage{subcaption}
\usepackage[pdftex]{graphicx}
\usepackage{cite}
\usepackage{upgreek}
\usepackage{cite}
\usepackage{xcolor,soul,framed} %,caption
\colorlet{shadecolor}{yellow}
\usepackage{url}
\usepackage{array}
\usepackage{caption}
\usepackage{balance}
\colorlet{shadecolor}{yellow}
\usepackage{color,soul}
\usepackage[cmex10]{amsmath}
\usepackage{algorithmic}
\usepackage{mdwmath}
\usepackage{mdwtab}
\usepackage{eqparbox}
\usepackage{array}
\usepackage{fixltx2e}
\usepackage{dblfloatfix}
\usepackage[mathscr]{euscript}
\usepackage{amsfonts}
\usepackage{amsmath}
\usepackage{commath}
\usepackage{upgreek}
\usepackage[long]{optidef}
\usepackage{bm}
\usepackage{algorithm}

\usepackage[open,openlevel=2]{bookmark}
\usepackage{cleveref}
\usepackage{textgreek}
\usepackage{booktabs,chemformula}
\usepackage{float}
\usepackage{amssymb}
\usepackage{bm}
\usepackage[thinlines]{easytable}
\usepackage{comment}
\usepackage{array, makecell}
\newcolumntype{P}[1]{>{\centering\arraybackslash}p{#1}}

\begin{document}
\DeclareGraphicsExtensions{.png}
%Note: Every paragraph is preceded by a comment in between two lines of asterisk signs to summarize the main message of this paragraph.  
%\title{Performance Analysis of long tail BSM latency in C-V2X Networks}
\IEEEoverridecommandlockouts
\title{Interleaved One-shot Semi-Persistent Scheduling for BSM Transmissions in C-V2X Networks}
\author{\IEEEauthorblockN{Abdurrahman Fouda$^{\dag}$, Randall~Berry$^{\dag}$
and Ivan~Vukovic$^{\ddag}$ 
\thanks{This project was supported in part by the Ford-Northwestern University Alliance.}
}
\IEEEauthorblockA{$^{\dag}$Department of Electrical and Computer Engineering, Northwestern University, Evanston, IL\\ 
$^{\ddag}$Ford Motor Company, Dearbon, MI\\
%Emails: 
abdurrahman.fouda@u.northwestern.edu, rberry@northwestern.edu, ivukovi6@ford.com}}

\markboth{}%
{Shell \MakeLowercase{\textit{et al.}}: Bare Demo of IEEEtran.cls for IEEE Journals}

\maketitle
\vspace{-0.5in}
\begin{abstract}
Cellular vehicle-to-everything (C-V2X) networks are regarded as one of the main pillars to enable efficient and sustainable Intelligent Transportation Systems (ITS) safety applications and services. Such services rely on the concept of exchanging periodic status updates (i.e., basic safety messages (BSMs)) between nearby vehicular users (VUEs). Hence, it is essential to ensure small inter-packet gaps (IPGs) between successive BSMs from nearby VUEs. Large IPGs, due to successive packet losses, can result in stale information at a VUE. In this paper, we study the tail behavior of the IPG and the information age (IA) distributions using C-V2X transmission mode 4 (a decentralized resource allocation method based on semi-persistent scheduling (SPS)). Specifically, we investigate improvements and trade-offs introduced by the SAE-specified concept of one-shot transmissions. We use a high-fidelity system-level simulator that closely follows the SPS process of C-V2X transmission mode 4 to evaluate the performance of the interleaved one-shot SPS transmissions. Our numerical results show that the tails of the IA and IPG complementary cumulative distribution functions (CCDFs) are significantly improved when one-shot transmissions are enabled in various simulation scenarios. 
\end{abstract}

\IEEEpeerreviewmaketitle

\section{Introduction}\label{sec_intro}
Vehicle-to-everything (V2X) communications have attracted a lot of academic and industrial attention over the last decade addressing improvements in vehicular systems’ reliability and efficiency. In its quest to address the needs of the automotive segment, the 3rd Generation Partnership Project (3GPP) has introduced support of C-V2X services (also referred to as LTE-V2X) in Releases 14 and 15~\cite{3gpp37885}. Further enhancements to support advanced new radio V2X (NR-V2X) were introduced in Release 16~\cite{nrpos}. It is expected that the support of advanced use cases for V2X applications will continue in Release 17 and beyond ~\cite{5gaa_slr}. Recently, cellular V2X (C-V2X) has emerged as a leading short-range radio technology for delivering V2X services in the 5.9 GHz Intelligent Transportation Systems (ITS) band. As part of Releases 14 and 15, 3GPP has introduced transmission modes 3 and 4 for C-V2X systems. In mode 3, base stations (eNBs) are responsible for the radio resource management (RRM) process. In mode 4, data packets and control information are exchanged directly between vehicular users (VUEs) without assistance from eNBs. In particular, VUEs use a distributed radio re source allocation scheme, namely, semi-persistent scheduling (SPS), to autonomously select their radio resources~\cite{3gpp36213}.

%and eNBs using the LTE PC5 sidelink. PC5 sidelink is a one-to-many communication interface that enables VUEs to communicate directly 

Given the real-time nature of ITS applications, it is vital to ensure the reliability and latency of such services. V2X direct communication is typically evaluated using two performance metrics, namely, information age (IA) and inter-packet gap (IPG), to evaluate the freshness of the last successfully received status update at destination VUEs. It has been observed through system simulation studies that the IPG (and IA) distribution in C-V2X systems can exhibit a long tail because the persistent nature of SPS can result in successive packet losses at the receiver VUEs~\cite{toyota,bspots}. SAE C-V2X technical Committee has specified a solution to remedy this by using one-shot transmissions to decrease the probability of persistent packet collisions which was not considered by previous studies. It is fair to assume that introducing another degree of randomness in the SPS scheduler such as one-shot transmissions may involve a trade-off with the average number of successfully received packets at destination VUEs. In this paper, we seek to study the impacts of using one-shot transmissions in C-V2X systems to improve the IPG and IA tail behavior and the trade-offs with average packet reception rate (PRR) through accurate discrete-event system simulations. 

SAE has defined basic safety messages (BSMs) for vehicular safety communications in V2X networks~\cite{J2735}. BSMs enable VUEs to share position, relative speed, and other mobility information with nearby VUEs. SAE is in the process of standardizing BSM transmissions using C-V2X for V2V scenarios~\cite{J3161}. C-V2X transmission mode 4 which does not require eNB support is a natural selection for vehicle-to-vehicle (V2V) communication and is being specified in~\cite{J3161}. C-V2X mode 4 defines V2V communication which is referred to in 3GPP terminology as sidelink (PC5 being the interface name). The sidelink performance of BSM transmission using mode 4 has been studied extensively in the literature to evaluate the reliability, in terms of PRR, and latency, in terms of IPG and/or IA, via: new collaboration methods between VUEs~\cite{piggyback}, enhanced SPS schemes~\cite{AugRA}, and by comparing DSRC (another vehicular communication protocol) with C-V2X via simulation campaigns~\cite{toyota}. Further, power optimization-based frameworks~\cite{ProbAoI1}, and AI-aware RRM schemes~\cite{RRM} have also been investigated to improve the IA tail (i.e., the worst-case scenarios) of V2X transmission mode 3. In~\cite{bspots}, the authors proposed an alternative method to decrease the probability of losing consecutive BSMs without impacting the transmission reliability by limiting the maximum duration, at which, a VUE keeps the same wireless resources for BSM transmission. However, none of the prior studies have discussed the tail improvement of the IPG distribution at destination VUEs using one-shot transmissions based on SAE draft specification~\cite{J3161}. Moreover, the results in [6] show a related tail improvement of the \textit{wireless blind spot duration} which measures the time a vehicle stays in a given choice of colliding resources. This is related to the IPG length but does not account for other causes of packet losses or the fact that vehicles can reselect into another collision, further extending the IPG.

This paper considers combining the concept of one-shot transmissions as specified in~\cite{J3161} with the sensing-based SPS to improve the IPG and IA tails of BSM transmissions using C-V2X mode 4. By randomly skipping SPS implicitly reserved virtual radio resource blocks (VRBs), one-shot transmissions mechanism seeks to decrease the probability of persistent BSM collisions at destination VUEs. We perform extensive simulation campaigns using a C++-based system simulator that closely follows the SPS process of C-V2X mode 4 specified in~\cite{3gpp36213} and the one-shot transmissions from~\cite{J3161} (formally known as event based transmissions). We evaluate the performance with and without one-shot transmissions at different bandwidth configurations, vehicle densities, and V2V distances in terms of IA, IPG, PRR, and channel busy ratio (CBR).

The numerical results show that one-shot transmissions can significantly improve the IA and IPG CCDF tails compared to the SPS implementation solely based on~\cite{3gpp36213}. We also demonstrate that the improvement in the 99.9th-percentile of the IA and IPG is robust against different simulation scenarios. Our results reveal that the IPG and IA tail improvements are influenced by the settings of the simulation scenarios such as the vehicle density, Tx-Rx separation, and bandwidth configuration. In particular, the best tail improvement can be achieved at the lowest vehicle density (125 VUE/km) compared with the higher vehicle densities of 400 and 800 VUE/km. Moreover, our results show that the closer the Tx-Rx pairs are, the better the tail improvement is. The numerical results also show that the expected degradation in the PRR performance (due to the interrupted persistency of SPS) is small compared to the significant improvements in the IA and IPG tails.

\section{Transmission Mode 4 in C-V2X Networks}\label{sec_tm4}

We consider a V2V broadcast scenario for BSM transmissions in a C-V2X system. Specifically, a set of VUEs utilizes C-V2X transmission mode 4 to periodically broadcast heartbeat BSMs to their neighbor vehicles in a half-duplex (HD) way. In doing so, VUEs transmit their data packets using the physical sidelink shared channel (PSSCH). The control packets are transmitted via the physical sidelink control channel (PSCCH) using the sidelink control information (SCI) format 1~\cite{3gpp36213}. In C-V2X, the physical channel is divided into sub-frames in the time domain and sub-channels in the frequency domain. The sub-frame width is 1 ms (i.e., a transmission time interval (TTI)), representing the time granularity for message scheduling in C-V2X transmission mode 4. The minimum allocation unit in the frequency domain for an LTE user is a physical resource block (PRB). A PRB spans 180 kHz in the frequency domain, 0.5 ms in the time domain, and contains 12 subcarriers separated by 15 kHz each. C-V2X defines a sub-channel as the minimum allocation unit to a VUE in the frequency domain. A sub-channel occupies a configurable number of PRBs, and the number of sub-channels per sub-frame is determined based on the operating bandwidth. 

In this paper, we refer to a sub-channel as a VRB and use the configuration of 10 PRBs per VRB. In C-V2X, data packets (e.g., BSMs) and SCI are transmitted in either adjacent or non-adjacent PRBs which are only required to be in the same sub-frame~\cite{3gpp36213}. For simplicity, we assume that the BSM PRBs and the corresponding SCI PRBs are always adjacent. Further, we use a fixed packet payload size for BSM and SCI, which occupies two contiguous VRBs per sub-frame~\cite{J3161}. The first two PRBs are reserved for SCI, and the remaining PRBs are reserved for data packets.

\subsection{Semi-persistent Scheduling}\label{sec_sub_sps}
In C-V2X transmission mode 4, VUEs utilize a sensing-based SPS scheduling scheme to autonomously allocate the radio resources without assistance from the cellular infrastructure. VUEs randomly select the required VRBs for BSM transmission from a candidate list of VRBs. The candidate list is defined using a pre-configured resource pool,  namely, the selection window. The selection window size in the time domain is given by $[n+T_{1},n+T_{2}]$. Here, $n$ denotes the BSM generation time, $T_{1}\le{4}$, and $T_{2}$ is determined based on the packet delay budget (PDB) where $T_{2}=\mathrm{max(PDB}-10,\,20\,\mathrm{ms})$~\cite{J3161}. The PDB is defined as the maximum allowed latency between the BSM generation time slot and the actual transmission time slot. It is worth mentioning that the BSM generation time here refers to the BSM arrival time from the application layer to the physical layer, at which time, the BSM becomes ready to be transmitted to neighboring VUEs. 

To facilitate SPS scheduling, SCIs are considered only if something is received and if so the average received reference signal resource power (RSRP) is used as a metric to exclude VRBs (from the selection window) whose RSRP is above a given threshold. Each VUE also excludes the VRBs which are used by other VUEs in the selection window. These VRBs are determined based on the received SCI in the last 1000 sub-frames. The total number of available VRBs for reselection should represent at least 20\% of all resources in the selection window resource pool. If not, a VUE keeps increasing the RSRP threshold by 3 dB iteratively until the 20\% target is met. The received signal strength indicator (RSSI) of the PSSCH VRBs is then used as a sorting metric to determine the best 20\% of the remaining VRBs in the selection window that experience the lowest received average RSSI. A VUE selects two contiguous VRBs of these resources randomly for the BSM transmission. For retransmissions, C-V2X mode 4 supports up to one redundant retransmission using hybrid automatic repeat request (HARQ) to improve the received signal-to-interference-plus-noise-ratio (SINR) and decrease consecutive packet losses~\cite{3gpp36321}. Each VUE selects VRBs for HARQ retransmissions from the same candidate list such that the sub-frame of the HARQ VRBs lies within 15 sub-frames from the sub-frame of the initially selected set of VRBs.

Once a VUE selects a new set of VRBs, it keeps \textit{reusing} them persistently for the next consecutive $C_{\mathrm{s}}$ BSM transmissions. Note that \textit{reusing} the same VRBs means using the same sub-frame in each selection window which is determined based on the BSM generation time $n$ and the PDB length. Once another BSM is generated at a given VUE, the VUE establishes the selection window and chooses the same sub-frame within that window for the next consecutive $C_{\mathrm{s}}$ BSM transmissions. Here, $C_{\mathrm{s}}$ is defined as the resource reselection counter (also referred to as the SPS interval) and is chosen uniformly at random between $[\alpha,\,\beta]$, where $\alpha$ and $\beta$ are fixed integers with $0<\alpha<\beta$. The resource reselection counter is decremented by one after each BSM transmission. When $C_{\mathrm{s}}$ reaches zero, a VUE reselects a new set of VRBs with a reselection probability $p_\mathrm{r}=1-p_\mathrm{k}$. Here, $p_\mathrm{k}$ is the probability to keep the current VRBs for the next BSM transmission after $C_{\mathrm{s}}$ reaches zero, where $p_\mathrm{k}\in[0,0.8]$ with a step of 0.2~\cite{J3161}. Once new VRBs are chosen or the current VRBs are kept, then a new SPS interval begins.

\subsection{One-shot transmissions}\label{sec_sub_1shot}
This section discusses how one-shot transmissions can be intertwined with SPS transmissions to improve the BSM tail behavior of C-V2X mode 4. Generally, one-shot transmissions rely on the idea of adding more randomness to the resource reselection process to avoid long IPGs (i.e., persistent packet collisions). Now, let $C_{\mathrm{o}}$ denote the one-shot resource reselection counter which is chosen uniformly at random between $[\rho,\,\sigma]$, where $\rho$ and $\sigma$ are fixed integers with $0<\rho<\sigma$. When one-shot transmissions are used, a VUE decrements both $C_{\mathrm{s}}$ and $C_{\mathrm{o}}$ by one every packet transmission. Next, we discuss how one-shot transmissions are implemented by considering the three possible scenarios of how $C_{\mathrm{s}}$ and $C_{\mathrm{o}}$ are related.   

First, when $C_{\mathrm{s}}$ reaches zero, while $C_{\mathrm{o}}>0$, the VUE again uses $p_{\mathrm{r}}$ to determine whether a new set of VRBs will be reselected or not. If a new set of VRBs is reselected, the VUE resets both counters and starts the process again. Otherwise, the VUE resets only the SPS counter $C_{\mathrm{s}}$ and decrements $C_{\mathrm{o}}$ by one. Second, when $C_{\mathrm{o}}$ reaches zero, a new set of resources is reselected and used only for the current transmission opportunity. The one-shot VRBs are selected using the same sensing-based reselection process discussed in Section~\ref{sec_sub_sps}. The VUE then resets $C_{\mathrm{o}}$ and uses the regular SPS-granted VRBs for the next BSM transmission opportunity. Finally, when both $C_{\mathrm{s}}$ and $C_{\mathrm{o}}$ reach zero simultaneously, the VUE rests both counters and uses $p_{\mathrm{r}}$ to determine whether a new set of VRBs will be reselected. If VUE decides to keep the old VRBs, it reselects a new set of one-shot VRBs and uses them for the current transmission opportunity. The old SPS-granted VRBs are then used for BSM transmission in the next transmission opportunity. If a new set of VRBs is reselected, VUE keeps using it for BSM transmission until either of $C_{\mathrm{s}}$, $C_{\mathrm{o}}$, or both of them expire and then repeats the above steps. 

\subsection{SAE Congestion Control}\label{sec_sub_cc}
In C-V2X scenarios, the packet loss rate increases when the number of competing vehicles over the same wireless resources increases. Hence, congestion control is vital to provide VUEs with adequate levels of successful packet reception. A decentralized congestion control mechanism for the application layer (which we use in this paper) is introduced in SAE J3161/1 to decrease the number of packet collisions in highly congested C-V2X scenarios~\cite{J3161}. In this approach, each VUE adjusts its BSM generation rate based on the estimated number of nearby transmitting vehicles. Let $w_{t}$ and $w_{k}$ denote two time intervals of sizes 1000 and 100 ms, respectively. Also, let $N^{v}_{\mathrm{c}}{(t)}$ denote the current number of unique neighbor vehicles in a given range of $r$ meters that $v^{\mathrm{th}}$ VUE has detected at least once in the previous 1000 ms. Here, $t$ is the last sub-frame of the time interval $w_{t}$. Also note that, a neighbor vehicle is determined to be unique if it has a unique ID in its BSM transmission. The average smoothed vehicle density at the $v^{\mathrm{th}}$ VUE can be calculated as: 
\begin{equation}
    N_{\mathrm{s}}^{v}(k)=\lambda\times{N^{v}_{\mathrm{c}}(t)}+(1-\lambda)\times{N^{v}_{\mathrm{s}}(k-1)}
\end{equation}
where $k$ and $k-1$ are the last sub-frames of the time intervals $w_{k}$ and $w_{k-1}$, respectively. Here, $\lambda$ is a weight factor between the current vehicle density and the previous estimated density. 

Now, let $I(k)$ represent the time interval between BSM packet generations in milliseconds for the $v^{\mathrm{th}}$ VUE at sub-frame $k$. SAE J3161/1 specifies that $I(k)$ be adapted based on the average estimated vehicle density as follows:   
\begin{equation}
        I(k)=
        \begin{cases}
            100 & \,\,\,N_{\mathrm{s}}^{v}(k) < B,\\
            100\times\frac{{N_{\mathrm{s}}^{v}(k)}}{B} & \,\,\,B \leq{N_{\mathrm{s}}^{v}(k)}<\frac{I_{\mathrm{max}}}{100}\times{B},\\
            I_{\mathrm{max}} &\,\,\, \frac{I_{\mathrm{max}}}{100}\times{B}\leq{N_{\mathrm{s}}^{v}(k)},
        \end{cases}
\end{equation}
where $B$ and $I_{\mathrm{max}}$ denote a pre-configured vehicle density co-efficient and the maximum allowed BSM packet generation interval, respectively. In other words, the generation of BSMs decreases linearly when the estimated vehicle density exceeds $B$ until it reaches a minimum value of $1/I_{\mathrm{max}}$. Note that $I(k)$ is updated every 100 ms regardless of the BSM generation rate. Further, we point out that $w_{t}$ and $w_{k}$ are not sliding time windows. Also, note that the above congestion control mechanism controls when BSMs are generated, once generated a BSM is then sent in the next sub-frame selected via SPS or one-shot reselection.  When congestion control is activated a VUE will not transmit in the selected sub-frame unless a new BSM has been generated.

\section{Numerical Analysis}\label{sec_results}
In this section, we illustrate the performance gains of using different one-shot configurations for BSM transmissions in C-V2X systems. We use a high-fidelity system-level C++ simulator that closely follows the SPS scheme for BSM transmission using C-V2X mode 4. Specifically, Monte Carlo simulations have been conducted to study the performance characteristics of different merit figures (e.g., IPG, IA, PRR, and CBR) for 10 MHz and 20 MHz bandwidths. In doing so, we use the ITU-R urban canyon path loss model in~\cite{toyota} with a Nakagami-$m$ distribution, which is given by:
\begin{equation}
    m(d)=
        \begin{cases}
            3 & \,\,\,d < 50,\\
            1.5 & \,\,\,50 \leq{d}<{150},\\
            1 &\,\,\, d\ge{150},
        \end{cases}
\end{equation}
where $m$ and $d$ denote the Nakagami fading parameter and the V2V distance in meters, respectively. Here, $\mathrm{Nakagami}(m,\,\Omega)$ random variates are generated using the $\mathrm{Gamma}(k,\,\theta)$ distribution as discussed in~\cite[Ch.~8]{SimulationModeling}, where $\Omega$ denotes the received non-faded power, $k=m $, and $\theta=\Omega/m$. 

VUEs are regularly spaced over a single lane in a highway scenario following the simulation settings in Table~\ref{tab_simpara}. Each vehicle is equipped with $n_t=1$ transmit and $n_r$ receive antennas and performs maximal ratio combining (MRC) on the received BSM~\cite[Ch.~3]{D.Tse}.  We use the in-band emission (IBE) models defined in\cite{3gpp37885} for 10 and 20 MHz bandwidths to account for the interference between different sub-channels in the same sub-frame. The link-level performance of the PSSCH and PSCCH is implemented using the  block error rate (BLER) versus SINR curves in~\cite{toyota,NIST}, respectively. A 3dB power boost is used to achieve an adequate performance of the PSCCH~\cite{3gpp36213}. We run the simulation scenarios for 500 seconds and collect the statistics only after the first 10 seconds, used as a simulation warm-up time to prevent any start-up bias. Numerical statistics are collected from vehicles located only in the middle third of the highway to minimize the boundary effects. Unless mentioned otherwise, we use the medium vehicle density (i.e., 400 VUE/km) to study the performance of the interleaved one-shot SPS without HARQ retransmissions.

\begin{table}
\renewcommand{\arraystretch}{1.3}
  \centering
  \caption{Simulation parameters.}\label{tab_simpara}
  \centering\renewcommand\cellalign{lc}
    \setcellgapes{500pt}\makegapedcells
        \begin{tabular}{l l}
        \hline
            \textbf{Parameter}  & \textbf{Value}\\
        \hline
            Scenario & Layout: 5 km single-lane Highway\\ 
            & Density: 125, 400 and 800 VUE/km\\
        \hline
            Channel model  & Path loss: ITU-R UHF urban canyon~\cite{toyota}\\  
            & In-band emission: 3GPP TR 36.885~\cite{3gpp37885}\\
            & Fast fading: Nakagami-$m$ distribution\\
        \hline
            Antenna settings & $n_t=1$, $n_r=2$, MRC receiver\\
        %\hline
        %   One-shot settings & $C_{\mathrm{o}}\in[2,\,6]$ and $[5,\,15]$\\
        \hline 
           SPS settings & $T_\mathrm{1}=4$, $T_\mathrm{2}=90$, $\mathrm{PDB}=100$ ms\\
           & $C_{\mathrm{s}}\in{[5,\,15]}$, $p_{\mathrm{k}}=0.8$\\
        \hline
            Resource pool settings & Carrier bandwidth: 10, 20 MHz at 5.9 GHz\\
            &Sub-channels per sub-frame: 5, 10\\
            & PRBs per sub-channel: 10\\ 
        \hline
            Power settings   & Tx power: 20 dBm, Noise figure: 6 dB\\
            & Thermal noise: -174 dBm / Hz\\ 
            & PSCCH power boost: 3dB \\
        \hline
            \makecell{10\% PER SINR\\threshold} & PSSCH: 3.7 dB, PSCCH: -1.3 dB\\
        \hline
            BSM packet size & Payload size: 300 bytes\\
            & Sub-channels per BSM: 2\\
            & PSSCH PRBs: 18, PSCCH PRBs: 2\\
        \hline
            CBR threshold  & -94 dBm\\
        %\hline
        %    HARQ settings  & Maximum retransmissions: 1\\ 
        %    & HARQ selection window: [-15, 15] ms\\ 
        \hline
            Congestion control & $d=100$ m, $I_{\mathrm{max}}=600$ ms,\\
            & $B=25$, $\lambda=0.05$\\%$l=100$ ms\\
        \hline
        %    Simulation time & 500 seconds\\
        %\hline
        \end{tabular}
        \vspace{-.15in}
\end{table}

\subsection{Inter-packet Gap (IPG)}\label{sec_sub_ipg}

% =======
% FIG. 01
% =======
\begin{figure*}
\subfloat[Distance bin 200 m]{\includegraphics[width=.33\textwidth]{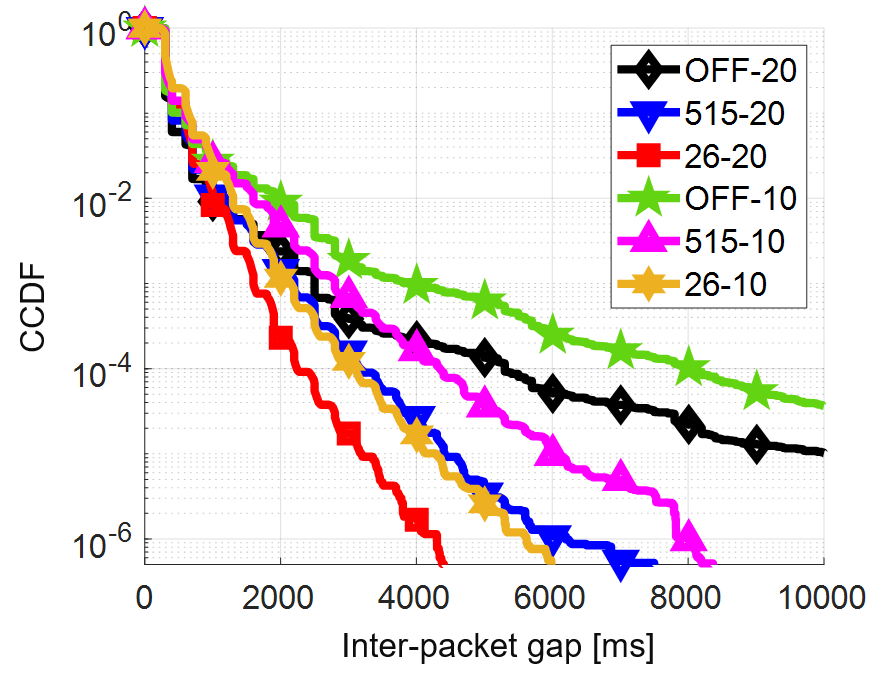}} 
\subfloat[Distance bin 300 m]{\includegraphics[width=.33\textwidth]{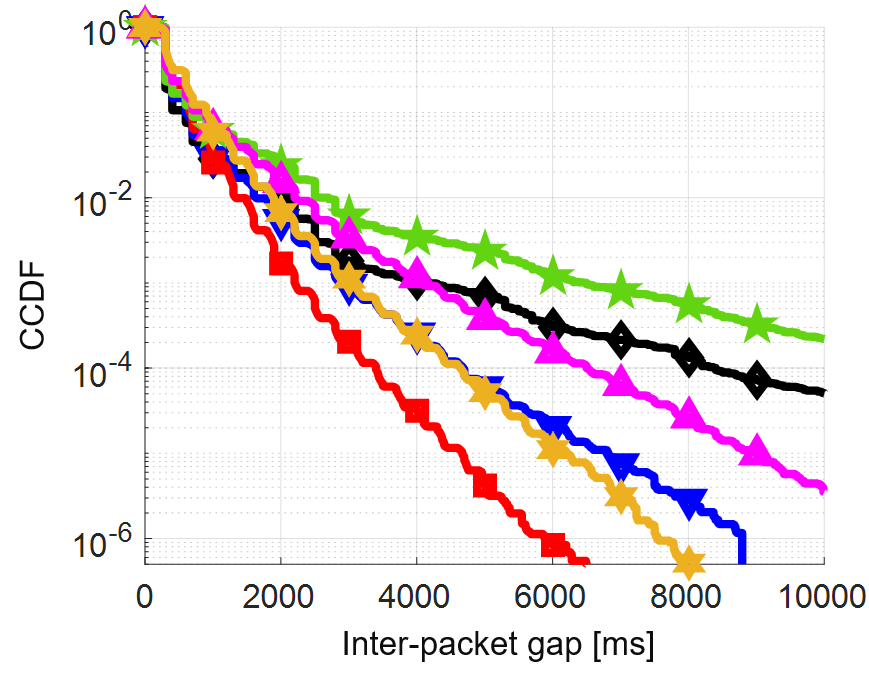}} 
\subfloat[Distance bin 400 m]{\includegraphics[width=.33\textwidth]{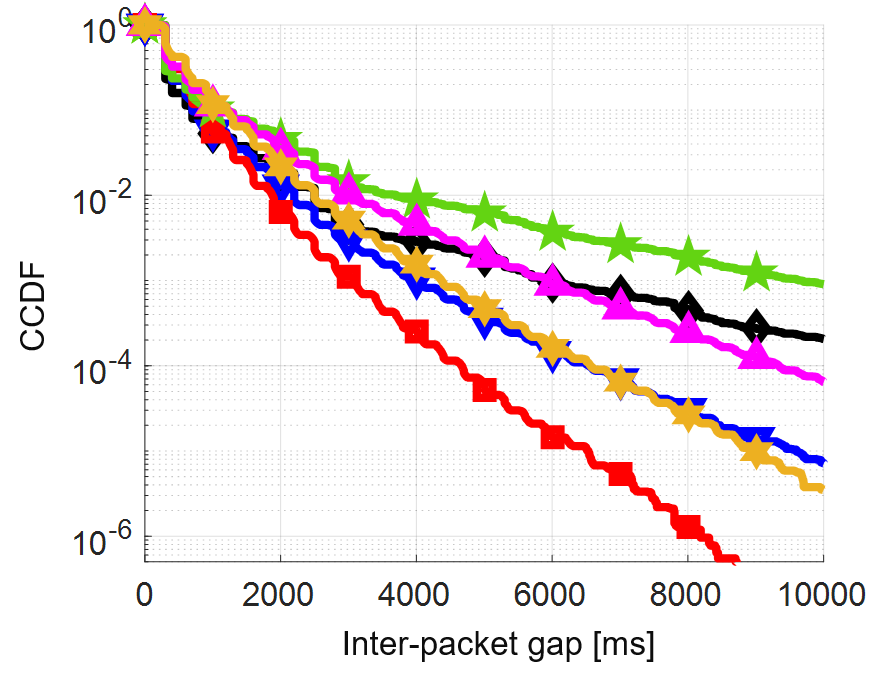}} 
\caption{CCDF of inter-packet gap.}
\vspace{-.25in}
\label{fig_ipgccdf}
\end{figure*}

% =======
% FIG. 02
% =======
\begin{figure}
  \begin{center}
  \includegraphics[width=6.5cm,height=6.5cm,,keepaspectratio]{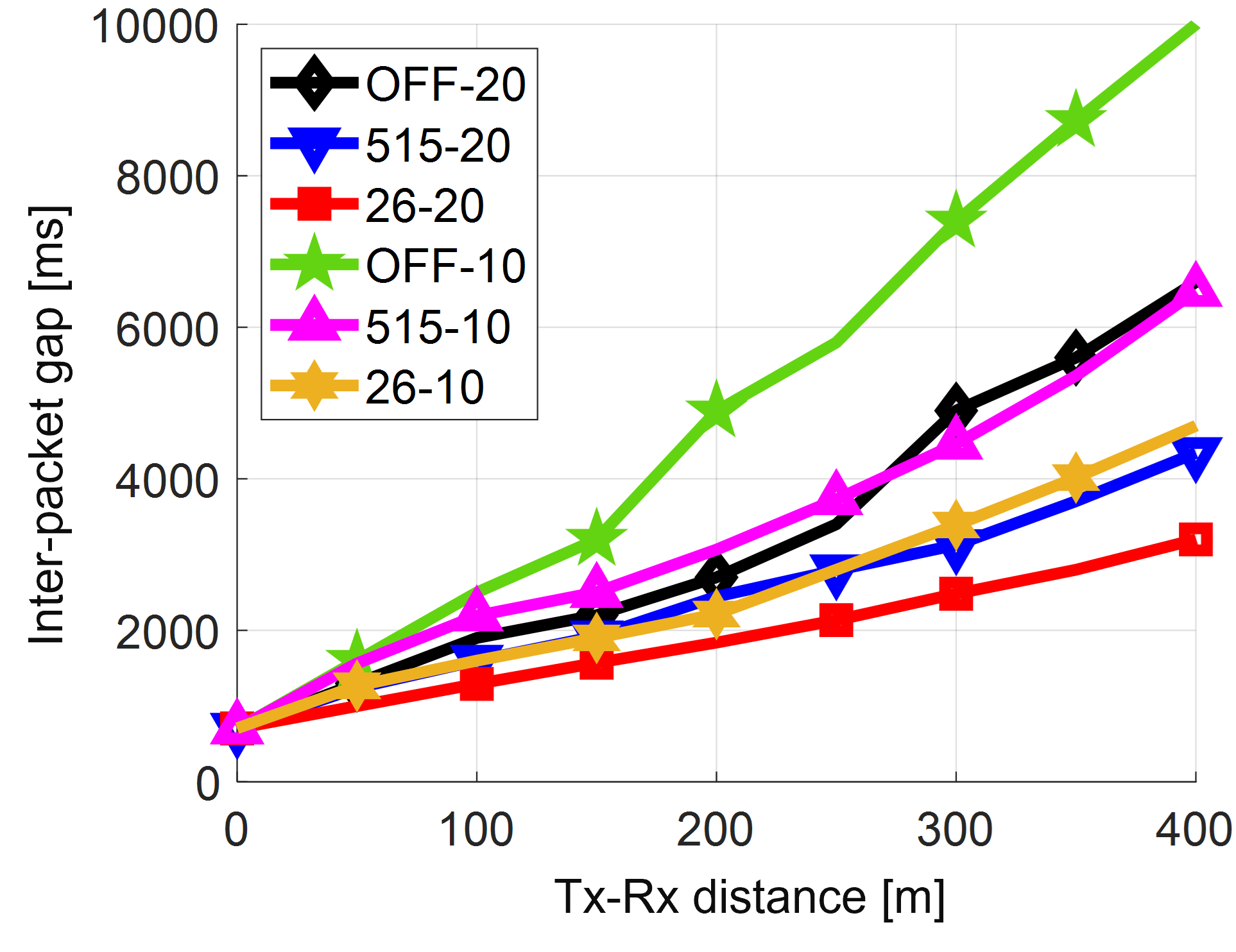}
  \caption{99.9th-percentile of inter-packet gap.}\label{fig_ipg999}
  \vspace{-.35in}
  \end{center}
\end{figure}
In this section, we evaluate the potential gains of using the one-shot transmission to improve the tail of IPG CCDF in C-V2X systems. For each pair of vehicles separated by distance $d$ meters, we measure the IPG as the time elapsed between each successive pair of successfully received BSMs. The IPG CCDF $F(i)$ is then given by the fraction of those values that exceed $i$ milliseconds. Fig.~\ref{fig_ipgccdf} shows the IPG CCDF with and without one-shot transmissions. The statistics are collected for 10 MHz and 20 MHz bandwidths at distance bins 200, 300, and 400 m. A V2V distance bin of $d$ represents all Tx-Rx pairs that are separated by $\zeta$ meters, where $\zeta\in{[d-25,\,d+25]}$. We use two configuration modes for one-shot transmissions, in which, $C_{\mathrm{o}}$ can be chosen uniformly at random between $[2,\,6]$ or $[5,\,15]$. The simulation scenarios' legends are denoted by $x$-$y$, where $x$ represents the $C_{\mathrm{o}}$ configuration and $y$ represents the operating bandwidth in MHz. As shown in Fig.~\ref{fig_ipgccdf}, with high probability, IPG is $\approx$ 310 ms (i.e., vehicles successfully receive BSMs every 310 ms). This is consistent with the message generation interval calculated using the SAE congestion control in Section~\ref{sec_sub_cc} for a vehicle density of 400 VUE/km.

From these figures, it can be seen that the one-shot transmissions do indeed improve the tail behavior of the IPG CCDF in all cases (i.e., they lead to a steeper drop in the tail). We measure this gain by calculating the average relative improvement in the IPG CCDF for all IPG values in the interval $[3,\,10]$ seconds. For example, let $F_{\mathrm{s}}(i)$ be the CCDF of the IPG evaluated at $i$ milliseconds of the simulation scenario $s$. The relative improvement in the IPG tail of simulation scenario \textit{26-20} is calculated as follows:
\begin{equation}
    \delta_{i}=\frac{F_{\mathrm{OFF-20}}(i)-F_{\mathrm{26-20}}(i)}{F_{\mathrm{OFF-20}}(i)}.
\end{equation}
The average relative improvement is then calculated as the average over vector $\Delta$ where $\Delta=[\delta_{i}: i\in{\mathcal{I}}]$, where $\mathcal{I}$ is a quantization of the interval $[3,10]$ seconds with steps of size 1 ms. As shown in Fig.~\ref{fig_ipgccdf}(a), at 20 MHz bandwidth, the one-shot transmissions with $C_{\mathrm{o}}$ configurations of $[5,\,15]$, and $[2,\,6]$ improve the IPG tail at distance bin 200 m by $\approx$ 94.7$\%$ and 99.6$\%$, respectively. The tail behavior improvement is slightly lower for 10 MHz bandwidth. Essentially, increasing the number of one-shot reselections ($[2,\,6]$ vs. $[5,\,15]$) decreases the number of instances with long IPGs (i.e., persistent BSM collisions) and improves the IPG tail behavior. 

Fig.~\ref{fig_ipgccdf} also demonstrates that the 20 MHz bandwidth configurations have better tail behavior than their counterparts with 10 MHz. This is because the higher number of available resources leads to a lower collision probability. %Table~\ref{tab_ipgstats} summarizes the relative IPG CCDF performance gains of using one-shot transmissions for different simulation scenarios. 
It is worth noting that, at V2V distance bin of 200 m, the IPG tail performance of the 10 MHz bandwidth with $C_{\mathrm{o}}\in[2,\,6]$ is better than that of the 20 MHz case without one-shot transmissions and with $C_{\mathrm{o}}\in[5,\,15]$. In other words, the negative impacts of using smaller bandwidths can be avoided by using rapid one-shot reselections (i.e., $[2,\,6]$ configuration). Fig.~\ref{fig_ipgccdf}(c) also reveals a smaller improvement in the IPG tail when the Tx-Rx pairs are further separated from each other (i.e., distance bin 400 m vs. 200 m). As the V2V separation between Tx-Rx pairs increases, the number of interferers increases, and IPG increases due to the higher number of persistent collisions. Hence, one-shot transmissions are less likely to stop long IPGs with a high number of interferers.
\begin{comment}
\begin{table}
\renewcommand{\arraystretch}{1.3}
    %\makegapedcells
  \centering
  \caption{Average relative gains in IPG CCDF.}\label{tab_ipgstats}
        \begin{tabular}{ P{1.8cm} P{.8cm} P{.8cm} P{.8cm} P{.8cm}}
        \hline
        \textbf{V2V distance}  & \textbf{515-20} & \textbf{26-20} & \textbf{515-10} & \textbf{26-10}\\ \hline
        200 & .94731 & .99638 & .93058 & .99298\\ \hline
        300 & .90240 & .98827 & .84946 & .97370\\ \hline
        400 & .82197 & .96504 & .73386 & .92858\\ \hline
        \end{tabular}
        \vspace{-.15in}
\end{table}
\end{comment}
% =======
% FIG. 03
% =======
\begin{figure*}
\subfloat[Distance bin 200 m]{\includegraphics[width=.33\textwidth]{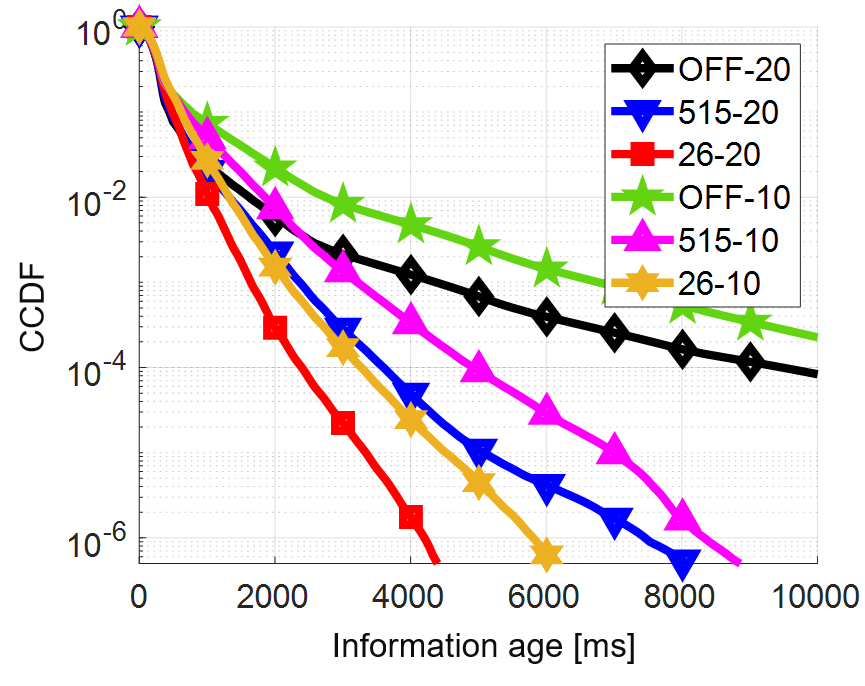}} 
\subfloat[Distance bin 300 m]{\includegraphics[width=.33\textwidth]{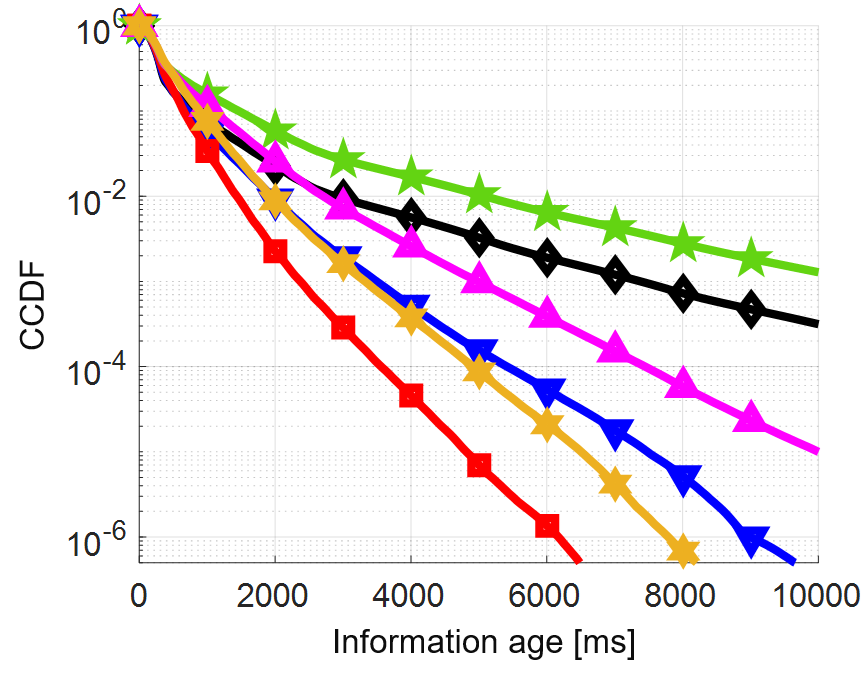}} 
\subfloat[Distance bin 400 m]{\includegraphics[width=.33\textwidth]{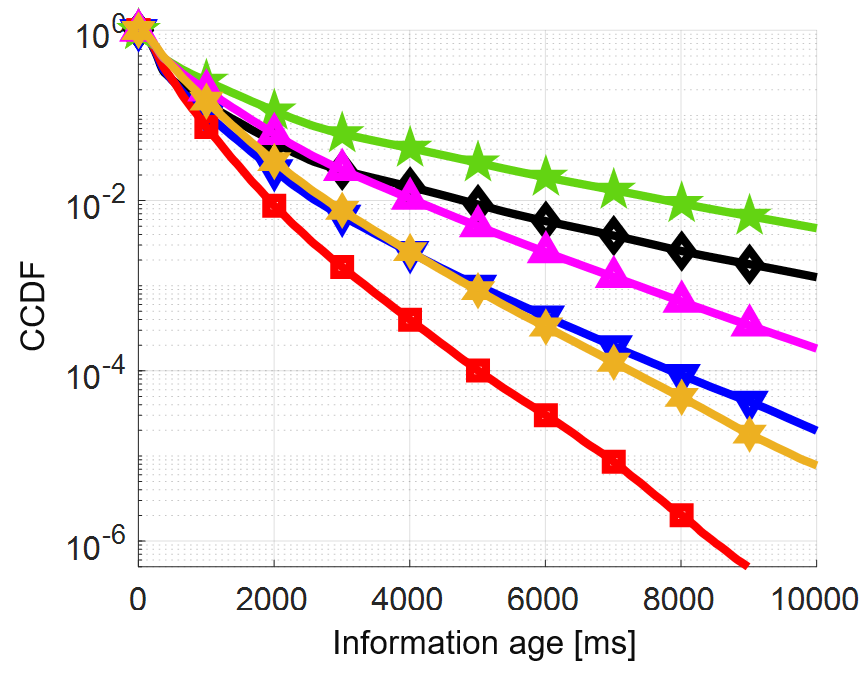}} 
\caption{CCDF of information age.}
\label{fig_iaccdf}
\vspace{-.25in}
\end{figure*}

% =======
% FIG. 04
% =======
\begin{figure}
  \begin{center}
  \includegraphics[width=6.5cm,height=6.5cm,,keepaspectratio]{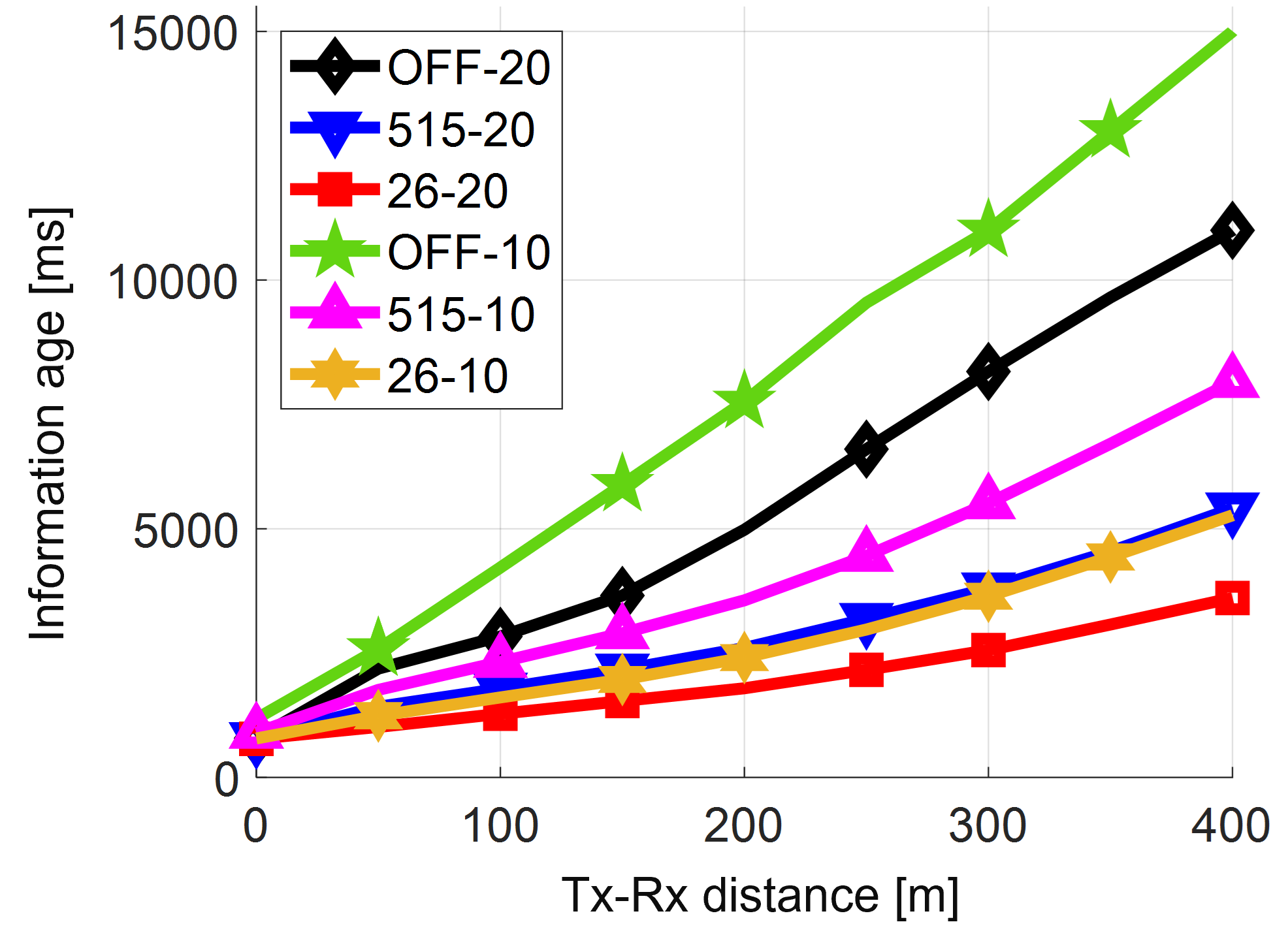}
  \caption{99.9th-percentile of information age.}\label{fig_ia999}
  \vspace{-.25in}
  \end{center}
\end{figure}

Fig.~\ref{fig_ipg999} provides another way to see the advantage of using the one-shot transmissions to improve the IPG tail in C-V2X networks. Specifically, we compare the length of IPGs that are larger than 99.9\% of the recorded IPGs at a given V2V distance. Again, each distance bin represents 50 m. We measure the improvement in the IPG 99.9th-percentile by comparing the IPG values of different simulation scenarios at a specific V2V distance bin. As shown in Fig.~\ref{fig_ipg999}, using $[5,\,15]$ and $[2,\,6]$ configurations of $C_{\mathrm{o}}$ with 20 MHz bandwidth improves the 99.9th-percentile of IPG by $\approx$ 19.6\%, and 39\% at distance bin 200 m, and by $\approx$ 34.5\% and 51.1\% at distance bin 300 m, respectively. A similar improvement is achieved when the same configurations are used with 10 MHz bandwidth. 

\subsection{Information Age (IA)}\label{sec_sub_ia}

Another important metric in V2X networks which is related to IPG is IA. This metric periodically measures the time elapsed since the generation of the latest successfully received BSM at destination VUEs. The IA of a BSM received at VUE B from VUE A at $t_{\mathrm{c}}$ is defined as follows:
\begin{equation}
    I_{\mathrm{B,A}}=t_{\mathrm{c}}-t_{\mathrm{s,B}}+\eta_{\mathrm{A}},
\end{equation}
where $t_{\mathrm{c}}$ and $t_{\mathrm{s, B}}$ denote the current time and the time of the last successfully received BSM at VUE B, respectively. $\eta_{\mathrm{A}}$ denotes the application-physical layer latency of VUE A, and is defined as $\eta_{\mathrm{A}}=t_{\mathrm{g,A}}-t_{\mathrm{r,A}}$, where $t_{\mathrm{g,A}}$ and $t_{\mathrm{r,A}}$ denote the BSM generation time and the actual transmission time at VUE A, respectively.  Here, we use a \textit{uniform} time sampling of the IA sawtooth sample path to collect the IA data based on time samples not the received BSMs (i.e., peak ages). In particular, the IA CCDF $F(i)$ can be interpreted as the fraction of time that the IA exceeds $i$ milliseconds. Fig.~\ref{fig_iaccdf}(a) shows that the $[5,\,15]$ and $[2,\,6]$ one-shot configurations with 20 MHz improve the IA tail by $\approx$ 98.1$\%$ and 99.9$\%$ at V2V distance bin 200 m, respectively, where the improvement is calculated using the same approach as we used for the IPG tail.
\begin{comment}
\begin{table}
\renewcommand{\arraystretch}{1.3}
    %\makegapedcells
  \centering
  \caption{Relative gains in IA statistics.}\label{tab_iastats}
        \begin{tabular}{ P{1.8cm} P{.8cm} P{.8cm} P{.8cm} P{.8cm}}
        \hline
        \textbf{V2V distance}  & \textbf{515-20} & \textbf{26-20} & \textbf{515-10} & \textbf{26-10}\\ \hline
        \multicolumn{5}{c}{\textbf{Average improvement of IA CCDF}}\\ \hline
        200 & .98139 & .99925 & .96970 & .99780\\ \hline
        300 & .95974 & .99670 & .92808 & .99139\\ \hline
        \end{tabular}
        \vspace{-.15in}
\end{table} 
\end{comment}

Fig.~\ref{fig_iaccdf} also reveals that these results are consistent with the results in Figs.~\ref{fig_ipgccdf} and~\ref{fig_ipg999}. Specifically, IA performance of the $[2,\,6]$ configuration at 10 MHz outperforms that of the 20 MHz without one-shot transmissions and with the $[5,\,15]$ configuration. Further, it demonstrates that using one-shot transmissions yields the best improvement in IA performance at distance bin 200 m (vs. all other distance bins). Fig.~\ref{fig_ia999} confirms the IA tail improvement when one-shot reselection is used. Specifically, $[2,\,6]$ configuration with 20 MHz improves the 99.9th-percentile of IA at distance bins 200 and 300 m by $\approx$ 64\% and 68.6\%, respectively. In summary, Figs.~\ref{fig_ipgccdf}-\ref{fig_ia999} demonstrate the advantages of using one-shot reselections to decrease the consecutive BSM losses at destination VUEs. The self-inherited random resource reselection nature of one-shot transmissions decreases the likelihood of multiple consecutive BSM losses caused by using SPS in C-V2X systems.
%It is worth mentioning that IA is one of the most important metrics to evaluate the freshness of the last successfully received status update at destination VUEs in V2X time-critical applications. 
%The IA tail improvement of different simulation scenarios is summarized in Table~\ref{tab_iastats}

\subsection{Packet Reception Ratio (PRR)}\label{sec_sub_prr}
% =======
% FIG. 05
% =======
\begin{figure}
  \begin{center}
  \includegraphics[width=6.5cm,height=6.5cm,,keepaspectratio]{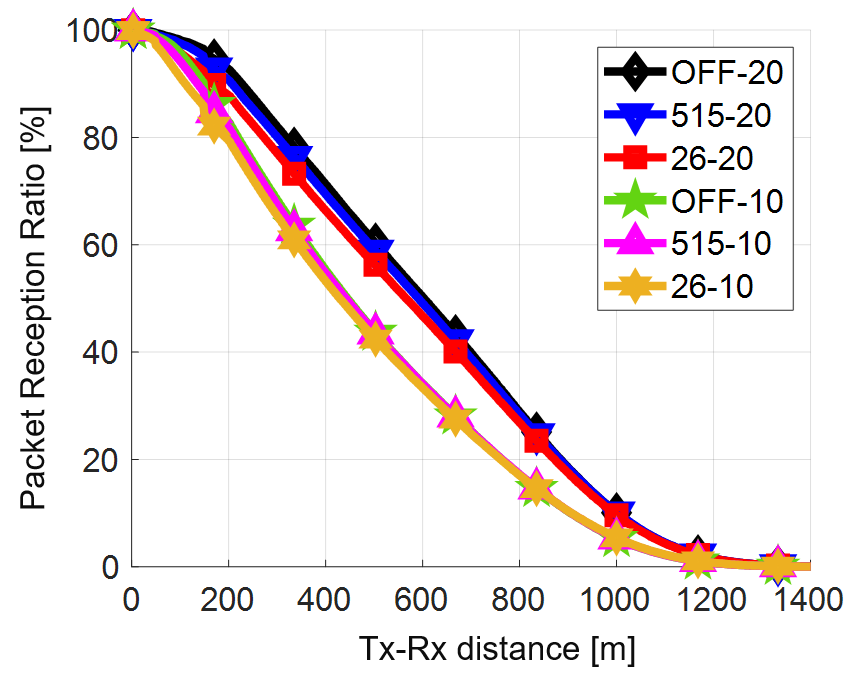}
  \caption{Packet reception ratio.}\label{fig_prr}
  \vspace{-.3in}
  \end{center}
\end{figure}
This section analyzes the PRR performance when one-shot reselection is used for BSM transmission. PRR is another instrumental performance metric used to evaluate the transmission reliability in C-V2X networks~\cite{3gpp37885}. In particular, PRR at a distance bin $d$ is calculated by $R/T$, where $T$ denotes the total number of transmitted BSMs between all Tx-Rx pairs that are separated by $d$ m, and $R$ is the number of successfully received BSMs among $T$. Fig.~\ref{fig_prr} shows the PRR with respect to the V2V distance where each distance bin represents 1 m (i.e., a one-to-one mapping). As expected, the one-shot configurations at 20 MHz outperform those at 10 MHz because of the higher number of available resources. Fig.~\ref{fig_prr} also reveals that the improvement in IPG and IA tails (when the one-shot transmissions are used) comes at the expense of slight degradation in PRR performance. Specifically, PRR performance degrades by $\approx$ 2.2\% and 5.3\% at distance bin 200 m when one-shot $[5,\,15]$ and $[2,\,6]$ configurations are used, respectively, with 20 MHz bandwidth. Similarly, PRR performance at the same distance bin drops by $\approx$ 1.4\% and 4.2\% when the same configurations are used with 10 MHz bandwidth. It is somewhat expected that since one-shot transmissions introduce more randomness into the SPS transmissions that they will lead to a lower PRR, which this result confirms. However, it appears that this degradation is relatively small and may be acceptable in order to get the improvement in the IPG and IA tails.

\subsection{PSSCH Channel Busy Ratio (CBR)}\label{sec_sub_cbr}
\begin{figure}
  \begin{center}
  \includegraphics[width=6.5cm,height=6.5cm,,keepaspectratio]{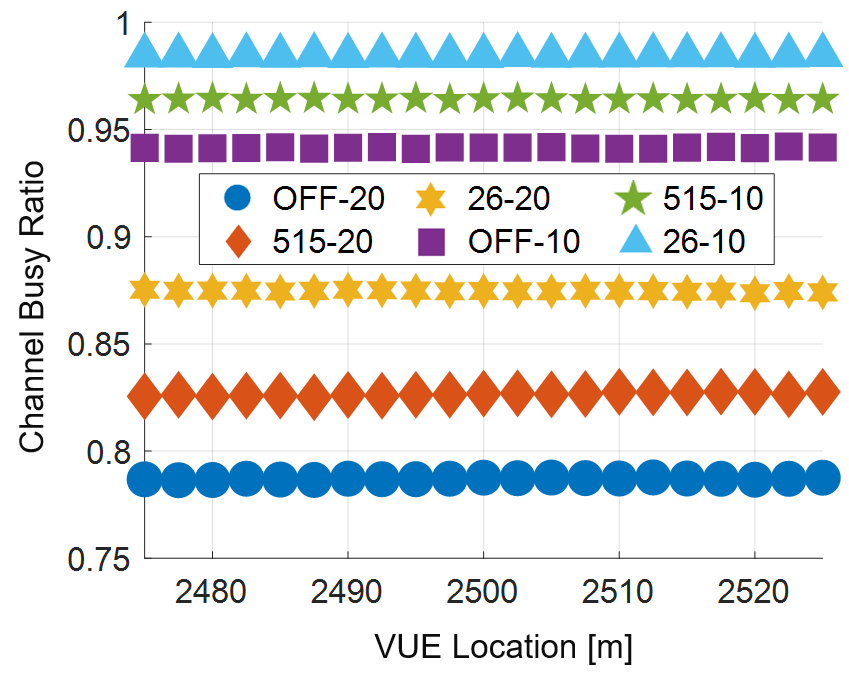}
  \caption{Channel busy ratio.}\label{fig_cbr}
  \vspace{-.25in}
  \end{center}
\end{figure}

We use CBR to evaluate the utilization of the  PSSCH wireless resources with different configurations of one-shot transmissions. Specifically, CBR is defined as $X/Y$, where $Y$ denotes the total number of VRBs in the SPS selection window (i.e., the most recent 100 ms), and $X$ denotes the VRBs \textit{among $Y$} whose average S-RSSI exceeds a threshold of -94 dBm at a given VUE~\cite{3gpp36321}. The average S-RSSI per VRB and the instantaneous CBR per VUE are updated periodically once per sub-frame (i.e., 1 ms) and measured over a sliding window of the last 1000 ms and 100 ms, respectively. Monte Carlo simulation is then used to calculate the average CBR over the simulation time. Fig.~\ref{fig_cbr} shows the average CBR of VUEs located within 25 m from the center of the highway. As shown in Fig.~\ref{fig_cbr}, CBR dropped from $\approx$ 94.2\% to 78.8\% when 20 MHz is used instead of 10 MHz without one-shot transmissions because of the higher number of available wireless resources for the same number of vehicles.  %For each VRB, the average S-RSSI is measured over a sliding window of the last 1000 ms and updated periodically every sub-frame (i.e., 1 ms). Similarly, the instantaneous CBR per VUE is updated periodically every 1 ms using a sliding window of the last 100 ms at every update.

Fig.~\ref{fig_cbr} also reveals that the $[5,\,15]$ one-shot configuration increases CBR from 94.2\% to 96.5\%, and from 78.8\% to 82.7\% at 10 and 20 MHz bandwidths, respectively. Similarly, the $[2,\,6]$ configuration increases CBR to 98.4\% and 87.5\% at 10 and 20 MHz, respectively. This is because the average S-RSSI per VRB is measured over a sliding window of the past 1000 sub-frames not only the current sub-frame. Essentially, using the one-shot transmissions decreases the resource reservation interval, during which a VUE keeps using the same VRBs for BSM transmission. Consequently, the total number of occupied VRBs (i.e., VRBs whose average S-RSSI exceeds -94 dBm) in the most recent 1000 ms increases as the number of one-shot reselections increases (see $[2,\,6]$ vs. $[5,\,15]$ configurations in Fig.~\ref{fig_cbr}). It is worth mentioning that the CBR statistics (with and without one-shot) could have been essentially the same if the S-RSSI per VRB was measured using only the most recent sub-frame (i.e., 1 ms). This is because the SPS-granted VRBs, of the current sub-frame, are considered free resources if a given VUE decides not to use them and does one-shot transmission using other VRBs.   

% =======
% FIG. 06
% =======

\subsection{Performance Evaluation of other densities}\label{sec_sub_otherd}
% =======
% FIG. 07
% =======
\begin{figure}
  \begin{center}
  \includegraphics[width=8.85cm,height=8.85cm,,keepaspectratio]{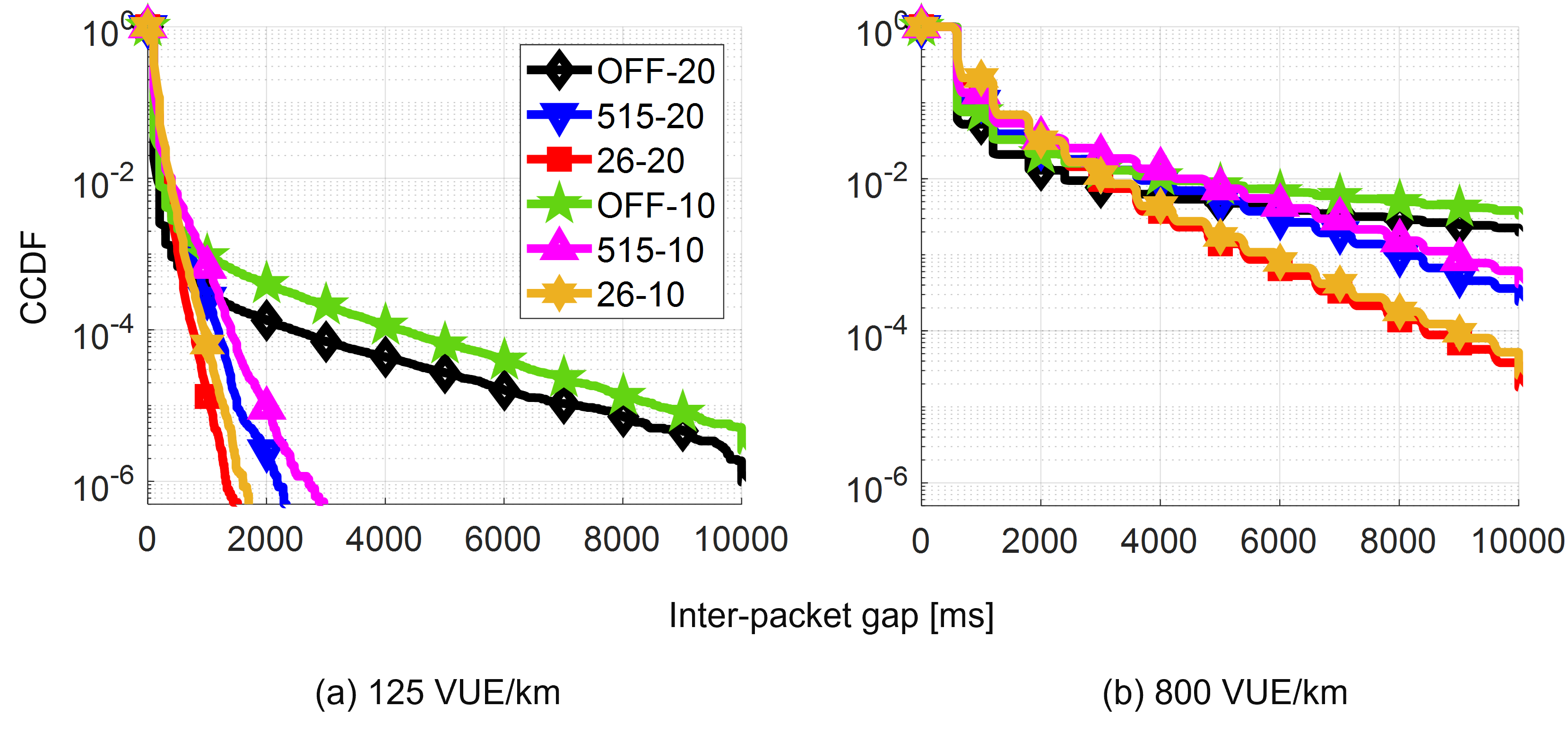}
  \caption{CCDF of inter-packet gaps for 200 m distance bin.}\label{fig_ipg6254000}
  \vspace{-.25in}
  \end{center}
\end{figure}

% =======
% FIG. 08
% =======
\begin{figure}
  \begin{center}
  \includegraphics[width=8.85cm,height=8.85cm,,keepaspectratio]{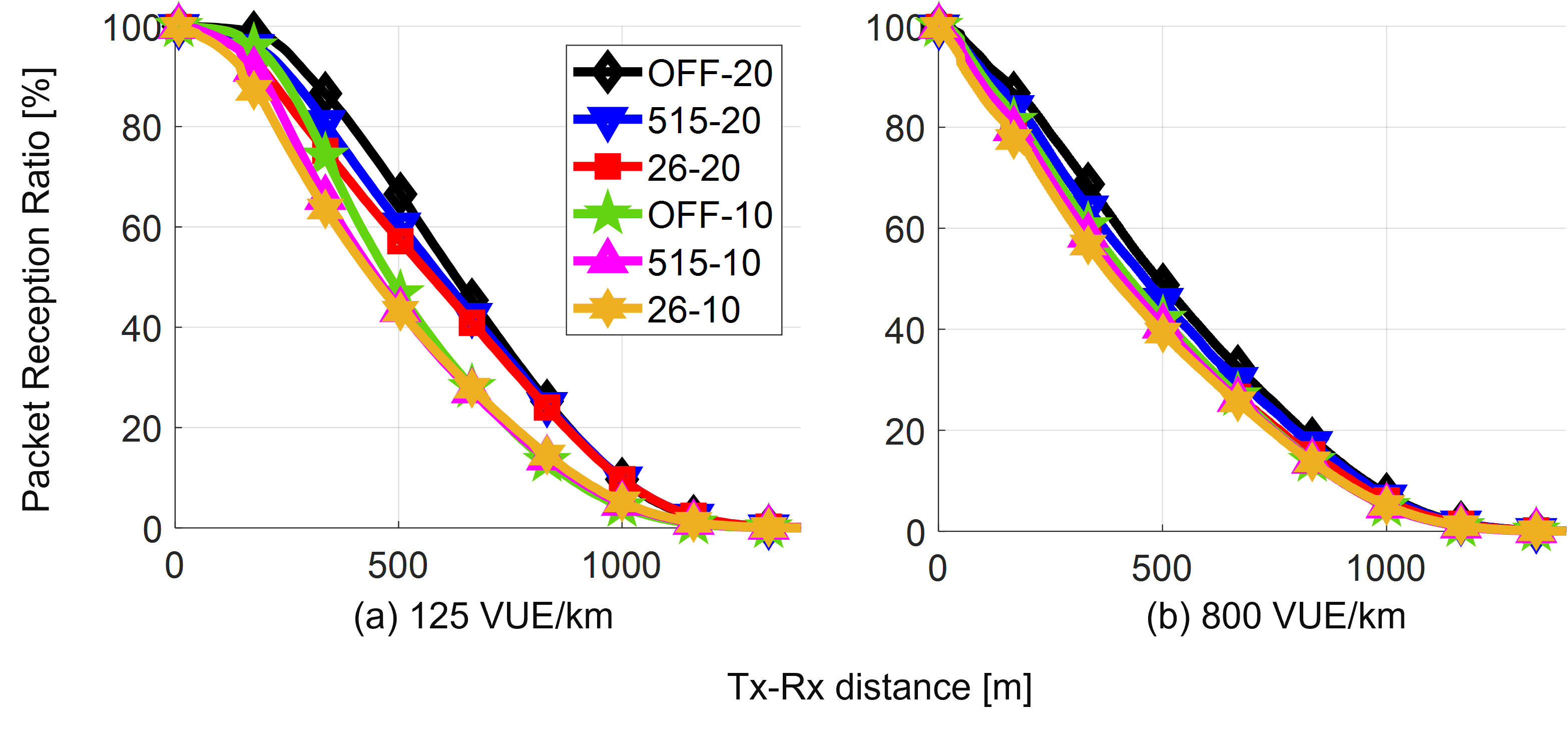}
  \caption{Packet reception ration versus distance.}\label{fig_prr625400}
  \vspace{-.25in}
  \end{center}
\end{figure}

In this section, we evaluate the performance of the IPG tail and PRR with other vehicle densities. Specifically, we use a low density of 125 VUE/km and a high density of 800 VUE/km which correspond to a BSM generation interval of 100 and 600 ms, respectively (see Section~\ref{sec_sub_cc}). Here, we calculate the performance gain/degradation using the same approaches as in Sections~\ref{sec_sub_ipg} and~\ref{sec_sub_prr}. Fig~\ref{fig_ipg6254000}(a) shows that the IPG tails of all scenarios with one-shot transmissions at the 200 m distance bin are improved by almost 100\% at the 125 VUE/km vehicle density. On the other hand, Fig.~\ref{fig_ipg6254000}(b) shows smaller gains when one-shot transmissions are used with the highest vehicle density (i.e., 800 VUE/km). Essentially, increasing the number of VUEs decreases the number of available VRBs and increases the collision probability. Hence, using the one-shot transmissions (to stop the persistent collisions of the SPS) at high densities may not lead to a significant IPG tail improvement because it is more likely that a given VUE will do a one-shot transmission (i.e., reselect) into another collision.

Fig.~\ref{fig_ipg6254000} also suggests that using the $[2,\,6]$ configuration of one-shot transmissions (at both bandwidths) with 800 VUE/km may slightly decrease the IPG tail performance for all IPGs $<\xi$, where $\xi=3$ seconds, because of the higher number of collisions. This threshold decreases as the vehicle density decreases because of the lower number of collisions. In particular, $\xi=1$ second and $500$ ms at 400 and 125 VUE/km, respectively. The same trend is observed for the $[5,\,15]$ one-shot configuration with slightly different thresholds. We point out that all above observations also apply for the other V2V distance bins and IA statistics. Fig.~\ref{fig_prr625400} shows that using the $[2,6]$ configuration of one-shot transmissions at 20 MHz bandwidth and 200 m V2V distance bin results in the highest PRR drop of $\approx$ 9\% at 800 VUE/km. The PRR drop is smaller at lower densities. In particular, using the same one-shot settings results in a PRR drop of $\approx$ 5\% and 7\% at 400 and 125 VUE/km, respectively. As expected, the highest PRR drop happens at the highest vehicle density when one-shot transmissions are used because of the smaller number of available VRBs.

\subsection{Performance Evaluation of HARQ retransmission}
% =======
% FIG. 09
% =======
\begin{figure*}
\subfloat[800 VUE/km] {\includegraphics[width=.33\textwidth]{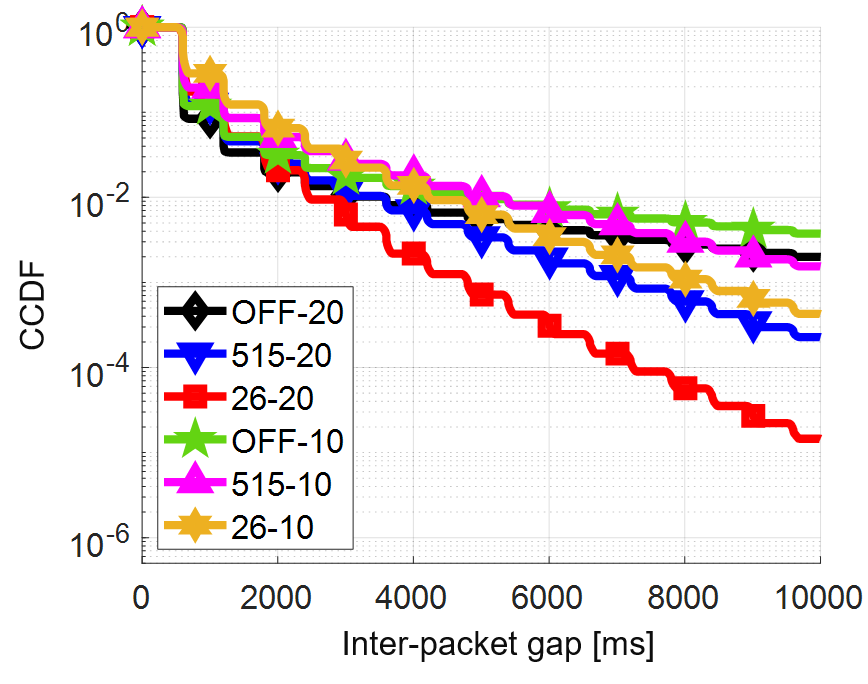}} 
\subfloat[400 VUE/km]
{\includegraphics[width=.33\textwidth]{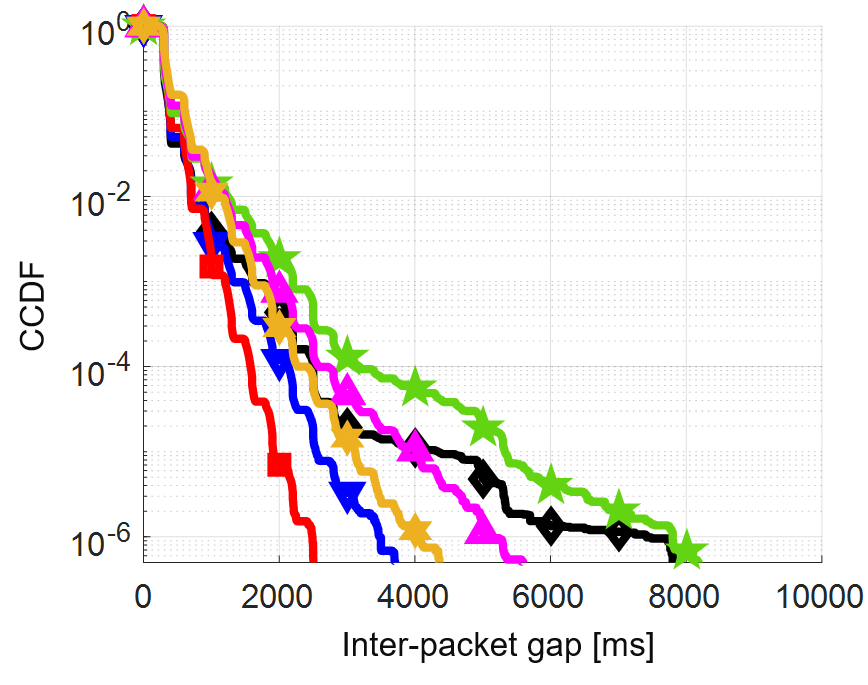}} 
\subfloat[125 VUE/km]
{\includegraphics[width=.33\textwidth]{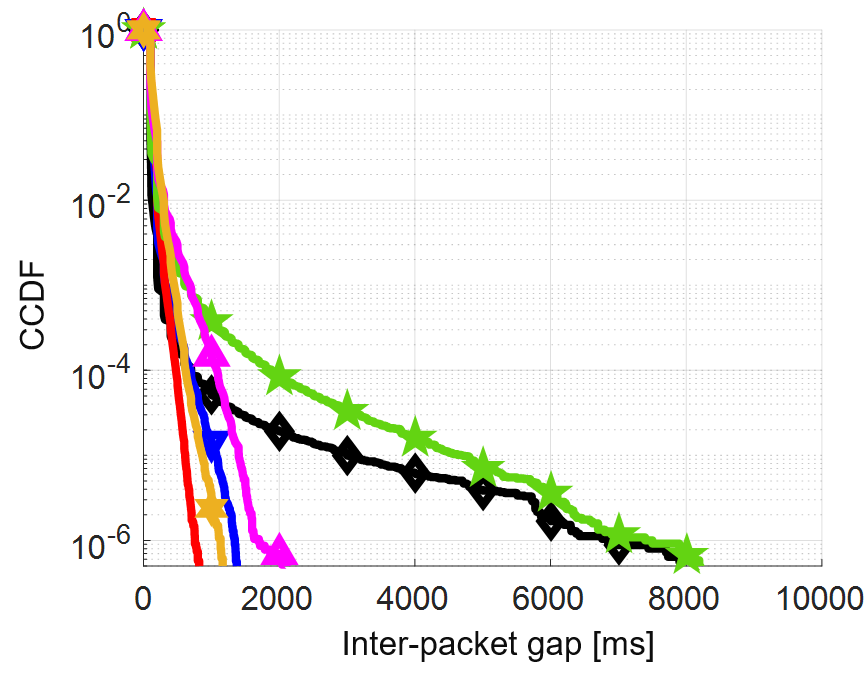}} 
\caption{CCDF of inter-packet gap for different vehicle densities using HARQ retransmission at 200 m distance bin.}
\vspace{-.15in}
\label{fig_ipgccdfh2}
\end{figure*}

% =======
% FIG. 10
% =======
\begin{figure*}
\subfloat[800 VUE/km] {\includegraphics[width=.33\textwidth]{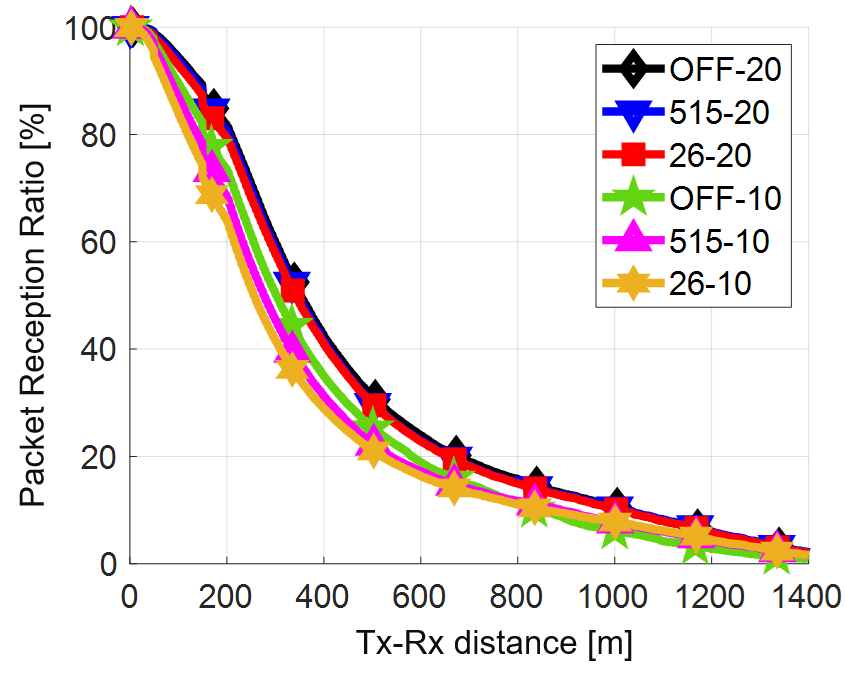}} 
\subfloat[400 VUE/km]
{\includegraphics[width=.33\textwidth]{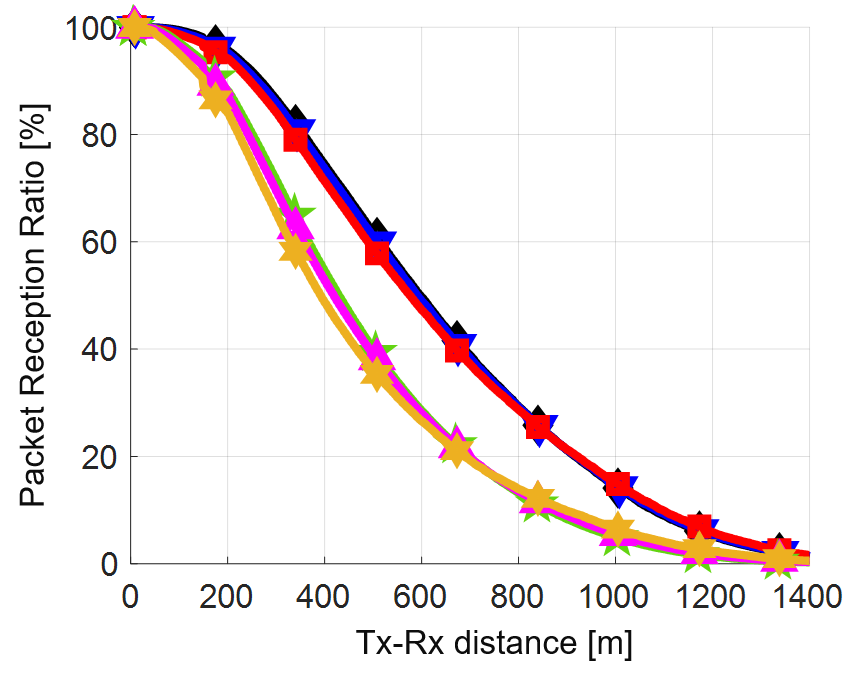}} 
\subfloat[125 VUE/km]
{\includegraphics[width=.33\textwidth]{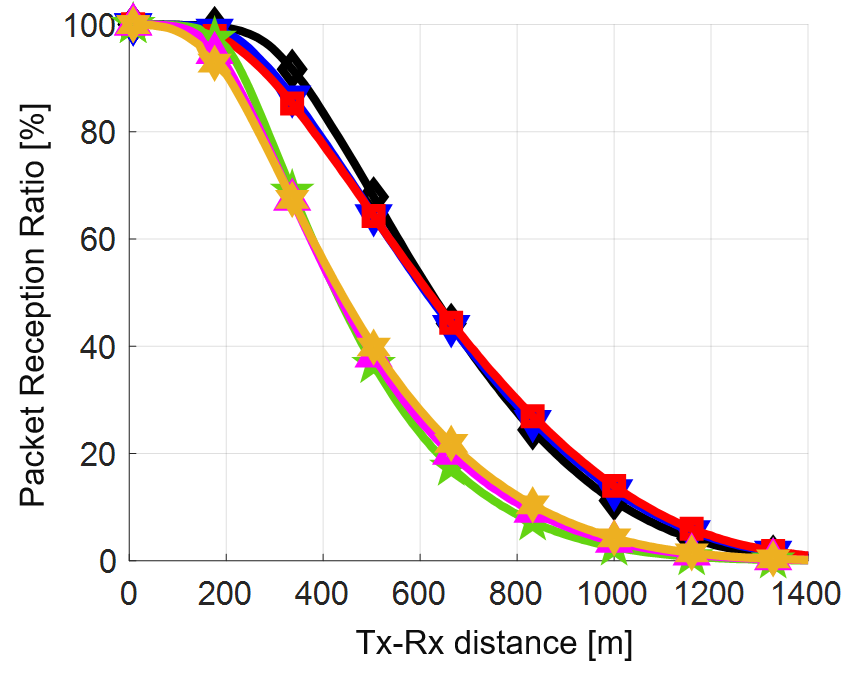}} 
\caption{Packet reception ratio for different vehicle densities using HARQ retransmission at 200 m distance bin.}
\vspace{-.2in}
\label{fig_prrh2}
\end{figure*}

In this section, we show the IPG tail improvement at the V2V distance bin of 200 m when HARQ retransmissions are used. As show in Fig.~\ref{fig_ipgccdfh2}(b), the one-shot configuration of $[2,\,6]$ improves the IPG tail by $\approx$ 99.9\% at 20 MHz with 400 VUE/km. A slightly smaller improvement ($\approx$ 95.9\%) is achieved at the 800 VUE/km vehicle density using the same simulation settings. Essentially, using HARQ retransmissions doubles the number of utilized VRBs for BSM transmissions and increases the congestion level. Again, using the one-shot transmission with highly congested scenarios (high vehicle density and HARQ retransmission) may decrease the IPG tail improvement because of the higher collision probability. This is more obvious at the 10 MHz bandwidth where the IPG tail is improved by only 30\% and 69\% using the $[5,\,15]$ and $[2,\,6]$ one-shot configurations, respectively, at 800 VUE/km. 

A similar trend is observed for the IA statistics. Further, the same observations of Sections~\ref{sec_sub_ipg} and~\ref{sec_sub_ia} apply for the other V2V distance bins. Fig.~\ref{fig_prrh2}(a) shows that the PRR performance drops by $\approx$ 2.7\% when the $[2,\,6]$ one-shot configuration is used at 20 MHz with HARQ retransmissions (compared with $\approx$ 9\% drop without HARQ, see Section~\ref{sec_sub_otherd}). Essentially, using HARQ retransmissions improves the received SINR and decreases the consecutive packet losses. In summary, Figs.~\ref{fig_ipgccdfh2} and~\ref{fig_prrh2} reveal that the higher resource consumption due to HARQ retransmissions does not jeopardize the IPG tail improvement. Further, it slightly decreases the PRR drop (caused by one-shot transmissions) by improving the received SINR at destination vehicles.  

\section{Conclusion}\label{sec_conc}
This paper investigates the impact of the SAE-standardized one-shot transmission feature to decrease the number of successive BSM losses at destination VUEs in C-V2X networks. In particular, we leverage the concept of one-shot transmissions which adds another degree of randomness in the SPS process to increase the probability of decoupling the transmissions of interfering VUEs from each other. This approach is evaluated via extensive discrete-event system simulation using a high-fidelity C++ simulator that closely follows the SPS process of C-V2X transmission mode 4. Our numerical analysis shows that using the interleaved one-shot SPS for BSM transmission significantly improves the IPG and IA CCDF tails in different C-V2X deployment scenarios, while only slightly decreasing the PRR. This trade-off becomes even more favorable for one-shot transmissions when HARQ retransmission is used. 
\balance
\bibliographystyle{IEEEtran}
\bibliography{IEEEabrv.bib,Bibliography.bib}

\begin{thebibliography}{10}
\providecommand{\url}[1]{#1}
\csname url@samestyle\endcsname
\providecommand{\newblock}{\relax}
\providecommand{\bibinfo}[2]{#2}
\providecommand{\BIBentrySTDinterwordspacing}{\spaceskip=0pt\relax}
\providecommand{\BIBentryALTinterwordstretchfactor}{4}
\providecommand{\BIBentryALTinterwordspacing}{\spaceskip=\fontdimen2\font plus
\BIBentryALTinterwordstretchfactor\fontdimen3\font minus
  \fontdimen4\font\relax}
\providecommand{\BIBforeignlanguage}[2]{{%
\expandafter\ifx\csname l@#1\endcsname\relax
\typeout{** WARNING: IEEEtran.bst: No hyphenation pattern has been}%
\typeout{** loaded for the language `#1'. Using the pattern for}%
\typeout{** the default language instead.}%
\else
\language=\csname l@#1\endcsname
\fi
#2}}
\providecommand{\BIBdecl}{\relax}
\BIBdecl

\bibitem{3gpp37885}
TSG{\,\,}RAN, ``Study on evaluation methodology of new ({V2X}) use cases for
  {LTE} and {NR},'' 3GPP TR37.885 v15.3.0, Jun. 2019.

\bibitem{nrpos}
A.~Fouda, R.~Keating, and A.~Ghosh, ``Dynamic selective positioning for
  high-precision accuracy in {5G NR V2X} networks,'' in \emph{Proc. {IEEE}
  Vehic. Technol.Conf. (VTC-Spring)}, Apr. 2021.

\bibitem{5gaa_slr}
5G{\,\,}Automotive{\,\,}Association, ``{C-V2X} use cases volume {II}: Examples
  and service level requirements,'' 5GAA, Oct. 2020.

\bibitem{3gpp36213}
TSG{\,\,}RAN, ``{LTE}; {E-UTRA}; physical layer procedures,'' 3GPP TS 36.213
  v16.4.0, Dec. 2020.

\bibitem{toyota}
T.~{Shimizu}, B.~{Cheng}, H.~{Lu}, and J.~{Kenney}, ``Comparative analysis of
  {DSRC} and {LTE-V2X PC5} mode 4 with {SAE} congestion control,'' in
  \emph{Proc. {IEEE} Vehic, Netw. Conf. (VNC)}, Dec. 2020, pp. 1--8.

\bibitem{bspots}
A.~Bazzi, C.~Campolo, A.~Molinaro, A.~O. Berthet, B.~M. Masini, and A.~Zanella,
  ``On wireless blind spots in the {C-V2X} sidelink,'' \emph{{IEEE} Trans. Veh.
  Technol.}, vol.~69, no.~8, pp. 9239--9243, Aug. 2020.

\bibitem{J2735}
SAE{\,\,}International, ``V2x communications message set dictionary,''
  J2735\_202007, Jul. 2020.

\bibitem{J3161}
------, ``On-board system requirements for {LTE-V2X V2V} safety
  communications,'' SAE-J3161/1 DRAFT, Jun. 2021.

\bibitem{piggyback}
F.~{Peng}, Z.~{Jiang}, S.~{Zhang}, and S.~{Xu}, ``Age of information optimized
  {MAC} in {V2X} sidelink via piggyback-based collaboration,'' \emph{{IEEE}
  Trans. Wireless Commun.}, vol.~20, no.~1, pp. 607--622, Jan. 2021.

\bibitem{AugRA}
Y.~{Jeon} and H.~{Kim}, ``An explicit reservation-augmented resource allocation
  scheme for {C-V2X} sidelink mode 4,'' \emph{{IEEE} Access}, vol.~8, pp.
  147\,241--147\,255, Aug. 2020.

\bibitem{ProbAoI1}
M.~K. {Abdel-Aziz}, S.~{Samarakoon}, C.~{Liu}, M.~{Bennis}, and W.~{Saad},
  ``Optimized age of information tail for ultra-reliable low-latency
  communications in vehicular networks,'' \emph{{IEEE} Trans. Commun.},
  vol.~68, no.~3, pp. 1911--1924, Mar. 2020.

\bibitem{RRM}
X.~{Chen}, C.~{Wu}, T.~{Chen}, H.~{Zhang}, Z.~{Liu}, Y.~{Zhang}, and
  M.~{Bennis}, ``Age of information aware radio resource management in
  vehicular networks: A proactive deep reinforcement learning perspective,''
  \emph{{IEEE} Trans. Wireless Commun.}, vol.~19, no.~4, pp. 2268--2281, Apr.
  2020.

\bibitem{3gpp36321}
TSG{\,\,}RAN, ``{LTE}; {E-UTRA}; medium control {(MAC)} protocol
  specification,'' 3GPP TS 36.321 v16.3.0, Dec. 2020.

\bibitem{SimulationModeling}
A.~M. Law, \emph{Simulation Modeling and Analysis}.\hskip 1em plus 0.5em minus
  0.4em\relax McGraw-Hill Education, 2015.

\bibitem{D.Tse}
D.~Tse and P.~Viswanath, \emph{Fundamentals of Wireless Communication}.\hskip
  1em plus 0.5em minus 0.4em\relax Cambridge Univ. Press, 2005.

\bibitem{NIST}
J.~Wang and R.~Rouil, ``{BLER} performance evaluation of {LTE} device-to-device
  communications,'' NISTIR 8157, Nov. 2016.

\end{thebibliography}
\end{document}